\documentclass[journal,letterpaper]{IEEEtran}
\usepackage[latin1]{inputenc}
\usepackage{amsmath,amsbsy,amssymb}
\usepackage{amsfonts,hyperref}
\usepackage{bbm}
\usepackage{verbatim}
\usepackage{amsthm}
\usepackage{units}
\usepackage{color}

\newcommand{\ket}[1]{|#1\rangle}
\newcommand{\bra}[1]{\langle #1|}
\newcommand{\bracket}[2]{\langle #1|#2\rangle}
\newcommand{\ketbra}[1]{|#1\rangle\langle #1|}

\newcommand{\eps}{\epsilon}
\newcommand{\tr}[1]{{\rm Tr}\left[#1\right]}
\newtheorem{theorem}{Theorem}
\newtheorem{lemma}{Lemma}

\newcommand{\hmax}[1]{{H}_{\max}^{#1}}
\newcommand{\hmin}[1]{{H}_{\min}^{#1}}

\newcommand{\lenc}[1]{\ell_{\rm enc}^{#1}}
\newcommand{\lext}[1]{\ell_{\rm ext}^{#1}}
\newcommand{\bpart}[2]{\{#1\}_{#2}}

\usepackage{amsmath}
\usepackage{amssymb}

\begin{document}
\author{Joseph M.~Renes and Renato Renner%
\thanks{J.M.\ Renes is with the Institut f\"ur Angewandte Physik, Technische Universit\"at Darmstadt, Hochschulstr. 4a, 64289 Darmstadt, Germany. Email: joe.renes@physik.tu-darmstadt.de.}
\thanks{R.\ Renner is with the Institute for Theoretical Physics, ETH Z\"urich, 8093 Z\"urich, Switzerland. E-mail: renner@phys.ethz.ch.}}

\title{One-Shot Classical Data Compression with Quantum Side Information and the Distillation of Common Randomness or Secret Keys}

\maketitle 

\begin{abstract}
The task of compressing classical information in the one-shot scenario is studied in the setting where the decompressor additionally has access to some given quantum side information. In this hybrid classical-quantum version of the famous Slepian-Wolf problem, the smooth max-entropy is found to govern the number of bits into which classical information can be compressed so that it can be reliably recovered from the compressed version and quantum side information. Combining this result with known results on privacy amplification then yields bounds on the amount of common randomness and secret key that can be recovered in one-shot from hybrid classical-quantum systems using one-way classical communication.
\end{abstract}
\begin{IEEEkeywords}
quantum information, data compression, Slepian-Wolf coding, smooth entropies
\end{IEEEkeywords}

\section{Introduction}


\IEEEPARstart{I}{nformation}
processing tasks, be they classical or quantum, are typically studied in the setting of asymptotically many independent and identically-distributed (i.i.d.) resources. Recent research has however extended our understanding to the \emph{one-shot} setting in which the resources are essentially arbitrary and structureless. Various protocols have been studied, such as extracting uniform randomness from a classical random variable, extracting randomness uncorrelated with possibly quantum adversaries (privacy amplification), as well as quantum data compression, state merging, entanglement distillation, and channel coding~(see \cite{renner_simple_2005} for an overview on classical protocols, \cite{dupuis_decoupling_2010} for an overview on quantum schemes based on decoupling, as well as~\cite{brandao_one-shot_2011,buscemi_distilling_2010,mosonyi_generalized_2009,wang_one-shot_2010,renes_noisy-channel_2011} for corresponding results related to entanglement manipulation and channel coding).  In this generalized setting, the Shannon or von Neumann entropies, which are normally used to quantify the strength of the available resources, need to be replaced by \emph{smooth entropies}, first introduced for the classical case in~\cite{renner_smooth_2004,renner_simple_2005}, and subsequently extended to the quantum case for the conditional and unconditional entropies in~\cite{renner_security_2005,renner_universally_2005} and the relative entropy in~\cite{datta_min-_2009}.

Here, we present one-shot results for the tasks of classical data compression with quantum side information and distillation of common randomness or shared secret keys using one-way communication.  Our results show that the relevant measure for characterizing the available resources is again the smooth entropy, in accordance with the aforementioned earlier findings. This confirms that, despite the generality of the one-shot approach, it is possible to formulate a variety of information-theoretic resource (in)equalities in terms of a single type of entropy measure.\footnote{Note that the smooth entropy comes in two versions, the smooth min-entropy and the smooth max-entropy. They are however dual to each other (see Eq.~\eqref{eq:dual} and subsequent discussion).} The situation is thus analogous to the standard i.i.d.-based theory, where the von Neumann entropy (which can be seen as a special case of the smooth entropy~\cite{tomamichel_fully_2009}) takes this role.

The problem of classical data compression with quantum side information at the decoder is a hybrid classical-quantum version of the famous Slepian-Wolf problem~\cite{slepian_noiseless_1973}, and was first studied in the asymptotic i.i.d.\ scenario by Winter~\cite{winter_coding_1999-1} and Devetak \& Winter~\cite{devetak_classical_2003}. There it is found that the classical random variable $X$ can be compressed at a rate given by the classical-quantum conditional entropy $H(X|B)=H(XB)-H(B)$ when the quantum system $B$ is available to the decoder. Here $H(\cdot)$ is the von Neumann entropy, and classical random variables are treated as quantum states diagonal in a fixed basis. We show that in the one-shot scenario the classical random variable can be compressed to a number of bits given by the smooth conditional max-entropy, and that this amount is optimal, up to small additive quantities involving the smoothing parameter. 

We then combine this result with known results on randomness extraction and privacy amplification to characterize protocols for both common randomness distillation and shared secret-key distillation from hybrid classical-quantum states in protocols using one way communication from the party holding the classical variable $X$ to the party holding the quantum system $B$. This task is relevant for post-processing in quantum key distribution protocols. Moreover, these two ``static'' problems are closely related to the ``dynamic'' tasks of transmitting classical information in public or private over a quantum channel. One shot results have been derived for the former case in~\cite{mosonyi_generalized_2009,wang_one-shot_2010}.
In~\cite{renes_noisy-channel_2011} we use the static protocols described here to directly construct optimal protocols for both public and private communication over quantum channels.


The paper is organized as follows. In the next section we describe the three tasks under consideration more concretely, give the definitions of smooth entropies as used here, and state our main results. The following section is then devoted to the proofs. Finally, we discuss some open questions and applications of this result.

\section{Definitions and Main Results}
Let us begin by describing the task of classical data compression with quantum side information at the decoder. 
Suppose that one party, Alice, holds a classical random variable $X$, while a different party, Bob, holds a quantum random variable, i.e.\ a quantum system $B$. The task of data compression with side information is for Alice to encode $X$ into another random variable $C$, such that Bob can reliably recover $X$ from $C$ and $B$. Clearly Alice could simply send $X$ itself, so we are interested in how small $C$ can be made in principle. We assume that the random variable $X$, as well as the state space of system $B$, are finite. The two random variables are defined by the ensemble $\{p_x,\varphi_x^B\}_{x\in\mathcal{X}}$, where $\mathcal{X}$ is the alphabet over which $X$ is defined, $p_x$ is its probability distribution, and $\varphi_x^B$ is the density operator of system $B$ when $X$ takes the value $x$.  In the following, we will describe this ensemble by  the classical-quantum (cq) state $\psi^{XB}=\sum_{x\in\mathcal{X}}p_x \ketbra{x}^X\otimes \varphi_x^B$; the compressed version of $X$ can be included by appending a system $C$.

A protocol is specified by the encoding map $\mathcal{E}:\mathcal{X}\rightarrow \{0,1\}^m$ and the decoding map $\mathcal{D}:S(\mathcal{H}^B)\times\{0,1\}^m\rightarrow \mathcal{X}$, where $m=\log_2|C|$ and $S(\mathcal{H}^B)$ is the set of density operators on the state space for system $B$.\footnote{Generally one may consider arbitrary sets for the output of the encoding map, not just those of size $2^m$. We do this here for simplicity.} (All logarithms are to be understood as base 2 in what follows.) The decoder generally consists of a quantum-mechanical measurement on system $B$, conditioned on the value of $C$. This takes the form of a POVM, a collection of positive operators $\Lambda_{x;c}^{B}$ such that $\sum_{x}\Lambda_{x;c}^{B}=\mathbbm{1}^{B}$ for all $c$. Therefore, the decoder is generally probabilistic. The protocol $(\mathcal{E},\mathcal{D})$ is said to be $\epsilon$-reliable or $\epsilon$-good when the average error probability is not greater than $\epsilon$:
\begin{align}
p_{\rm err}=1-\sum_{x\in\mathcal{X}}p_x \tr{\Lambda^{B}_{x;\mathcal{E}(x)}\varphi_x^B}\leq \epsilon.
\end{align}
Note that if we call $X'$ the output of the decoder, the probability of error is equal to the variational distance (the trace distance of classical random variables) of $p_{x,x'}$ to the ideal output $p_x\delta_{x,x'}$: $p_{\rm err}=\tfrac{1}{2}\sum_{x,x'}|p_x\delta_{x,x'}-p_{x,x'}|$. 
Finally, we denote by $\lenc{\epsilon}(X|B)_\psi$ the smallest achievable size of $\log|C|$ for an $\epsilon$-good protocol applied to the state $\psi^{XB}$. 

The two tasks distilling either common randomness or a shared secret key are very much related to the data compression problem. The goal now is for Alice and Bob to not only end up each holding a copy of the random variable $X$, but to further transform this into either shared uniform randomness or uniform randomness uncorrelated with a system $E$ held by an eavesdropper. Again information $C$ is sent from Alice to Bob, who each then generate new classical random variables $K_A$ and $K_B$ such that $K_A=K_B$ from $X$ and $(B,C)$, respectively. In each case we also demand that these outputs are uncorrelated with $C$. Common randomness distillation is a special case of secret key distillation with trivial $E$, so we focus on secret key distillation in what follows.

The quality of the output can be measured by the trace distance to the ideal state. The trace distance $D(\rho,\sigma)$ for two states $\rho$ and $\sigma$ is defined by $D(\rho,\sigma)\equiv\frac{1}{2}\|\rho-\sigma\|_1$, where $\|A\|_1\equiv\tr{\sqrt{A^\dagger A}}$ for arbitrary $A$. The output pair $K_A$ and $K_B$ of a protocol exchanging information $C$ is called an $\epsilon$-good secret key against $E$ if the output state $\rho^{K_AK_BCE}$ is such that $D(\rho^{K_AK_BCE},\kappa^{K_AK_B}\otimes\rho^{CE})\leq \epsilon$, where $\kappa^{K_AK_B}=\frac{1}{|K|}\sum_r \ketbra{k}^{K_A}\otimes\ketbra{k}^{K_B}$. The number of $\epsilon$-good common random bits that can be distilled in this manner starting from a shared state $\psi^{XB}$ we call $\ell_{\rm secr}^\epsilon(X;B|E)_\psi$. 

Our goal is to bound the quantities $\lenc{\epsilon}(X|B)_\psi$ 
and  $\ell_{\rm secr}^\epsilon(X;B|E)_\psi$ in terms of the smooth min- and max-entropies. First, the conditional max entropy for a state $\rho^{AB}$ is defined by 
\begin{align}
\label{eq:max}
\hmax{}(A|B)_\rho\equiv\max_{\sigma^B}\,\, 2\log F(\rho^{AB},\mathbbm{1}^A\otimes\sigma^B),
\end{align}
where the maximization is over positive, normalized states $\sigma$ and $F(\rho,\sigma)\equiv\|\sqrt{\rho}\sqrt{\sigma}\|_1$ is the fidelity of $\rho$ and $\sigma$. Dual to the conditional max-entropy is the conditional min-entropy,
\begin{align}
\label{eq:hmin}
\hmin{}(A|B)_\rho &\equiv \max_{\sigma^B}\left(-\log \lambda_{\min}(\rho^{AB},\sigma^{B})\right),
\end{align}
with $\lambda_{\min}(\rho^{AB},\sigma^B){\equiv}\min\left\{\lambda:\rho^{AB}\leq \lambda\mathbbm{1}^A\otimes \sigma^B\right\}$. The two are dual in the sense that
\begin{align} \label{eq:dual}
  \hmax{}(A|B)_{\rho}=-\hmin{}(A|C)_{\rho}
\end{align}
for $\rho^{ABC}$ a pure state~\cite{koenig_operational_2009}.
 
Each of these entropies can be \emph{smoothed} by considering possibly subnormalized states $\bar{\rho}^{AB}$ in the $\epsilon$-neighborhood of $\rho^{AB}$, defined using the purification distance 
$P(\rho,\sigma)\equiv\sqrt{1-F(\rho,\sigma)^2}$,
\begin{align}
B_\epsilon(\rho)\equiv\{\bar{\rho}:P(\rho,\bar{\rho})\leq \epsilon\}.
\end{align}  
Note that the purification distance is essentially equivalent to the trace distance, due to the bounds 
$D(\rho,\sigma)\leq P(\rho,\sigma)\leq\sqrt{2D(\rho,\sigma)}$~\cite{tomamichel_duality_2010}.
The smoothed entropies are then given by 
\begin{align}
\hmin{\epsilon}(A|B)_\rho&\equiv\max_{\bar{\rho}\in B_\epsilon(\rho^{AB})} \hmin{}(A|B)_{\bar{\rho}},\\
\label{eq:hmaxsmooth}
\hmax{\epsilon}(A|B)_\rho&\equiv\min_{\bar{\rho}\in B_\epsilon(\rho^{AB})} \hmax{}(A|B)_{\bar{\rho}}.
\end{align}
Furthermore, the dual of $\hmax{\epsilon}(A|B)_\rho$ is $\hmin{\epsilon}(A|C)_\rho$, so that taking the dual and smoothing can be performed in either order~\cite{tomamichel_duality_2010}.  

Now we can state our main results. 
\begin{theorem}[Classical Data Compression with Quantum Side Information at the Decoder]
\label{thm:csi}
Given any $\epsilon\geq 0$ and state $\psi^{XB}=\sum_x p_x \ketbra{x}^X\otimes \varphi_x^B$,
\begin{align*}
\lenc{\epsilon}(X|B)_\psi&\leq \hmax{{\epsilon_1}}(X|B)_\psi+2\log\tfrac{1}{\epsilon_2}+4,\\
\lenc{\epsilon}(X|B)_\psi&\geq \hmax{\sqrt{2\epsilon}}(X|B)_\psi,
\end{align*}
for $\epsilon_1,\epsilon_2\geq 0$ such that $\epsilon=\epsilon_1{+}\epsilon_2$.
\end{theorem}
\begin{theorem}[Secret Key Distillation with One-Way Public Communication]
\label{thm:skd}
Given any $\epsilon\geq 0$ and a state $\psi^{XBE}=\sum_x p_x \ketbra{x}^X\otimes \varphi_x^{BE}$ and $\epsilon{=}\epsilon_1{+}\epsilon_2$, $\epsilon'{=}\epsilon'_1{+}\epsilon_2$,
\begin{align*}
\ell_{\rm secr}^{\epsilon+\epsilon'}(X{;}B|E)_\psi&\geq \!\!\sup_{(U,V)\leftarrow X}\!\!\hmin{{\epsilon'_1}}(U|EV)_\psi\!-\!\hmax{{\epsilon_1}}(U|BV)_\psi\\
&\phantom{\geq \sup_{(U,V)\leftarrow X}}\,-4\log\tfrac{1}{\epsilon_2}-3\\
\ell_{\rm secr}^\epsilon(X{;}B|E)_\psi &\leq \!\!\sup_{(U,V)\leftarrow X}\!\!\hmin{\sqrt{2\epsilon}}(U|EV)_\psi\!-\!\hmax{\sqrt{2\epsilon}}(U|BV)_\psi.
\end{align*}
\end{theorem}

\section{Proof of the Main Results}
\label{sec:proof}
\subsection{Data Compression with Quantum Side Information}

To prove the lower bound of Theorem~\ref{thm:csi}, often called the direct part, we exhibit a protocol achieving it. The idea is for Alice to sufficiently narrow the set of states $\varphi_x^B$ which Bob is attempting to distinguish by providing him with the information $\mathcal{E}(x)$. Our protocol makes use of 2-universal hashing by the encoder and a variant of the pretty good measurement~\cite{hausladen_`pretty_1994} by the decoder. 
A \emph{family of hash functions} $f:\mathcal{X}\rightarrow \{0,1\}^m$ is called 2-universal if, when choosing a function randomly from this family, the probability of collision, $f(x)=f(y)$ for $x\neq y$, is at most the same as for random functions: Pr$_f[f(x)=f(y)]\leq 1/2^m$~\cite{carter_universal_1979}. 
The proof proceeds on the basis of the following two lemmas. The first is a bound on the error probability based on Lemma 2 of~\cite{hayashi_general_2003}.

\begin{lemma}
\label{lem:unravel}
Let $\psi^{XB}=\sum_{x\in\mathcal{X}} p_x\ketbra{x}^X\otimes \varphi_x^B$ be an arbitrary cq state with $\varphi^B=\sum_x p_x\varphi_x^B$, $F$ be a 2-universal family of hash functions $f:\mathcal{X}\rightarrow\{0,1\}^m$, and $P^{XB}$ be an operator of the form $P^{XB}=\sum_{x\in\mathcal{X}} \ketbra{x}^{X}\otimes \Pi_x^B$ with $0\leq \Pi_x\leq \mathbbm{1}$ for all $x\in\mathcal{X}$. Then there exists a family of measurements on $B$ indexed by $f\in F$ and $c\in\{0,1\}^m$ and having elements $\Lambda_{x;c,f}^B$ corresponding to outcomes $x$, such that 
$\Lambda_{x';c,f}=0$ when $f(x')\neq c$, and for which the error probability $\overline{p}_{\rm err}$ averaged over a random choice of $f\in F$ obeys
\begin{align*}
\overline{p}_{\rm err}&=\tfrac{1}{|F|}\sum_{f,x}p_x\tr{\left(\mathbbm{1}-\Lambda_{x;f(x),f}^B\right)\varphi_x^B}\\
&\leq  2\tr{(\mathbbm{1}{-}P^{XB})\psi^{XB}}{+}{4}\cdot{2^{-m}}\tr{P^{XB}(\mathbbm{1}^X{\otimes} \varphi^B)}.
\end{align*}

\end{lemma}
\begin{IEEEproof}
The measurement on $B$ is defined by the pretty-good measurement using all the $\Pi_x$ such that $f(x)=c$. It has elements
\begin{align*}
\Lambda_{x;c,f}=\left(\sum_{x':f(x')=c}\Pi_{x'}\right)^{\!\!\!\!-\frac{1}{2}}\Pi_x\left(\sum_{x':f(x')=c}\Pi_{x'}\right)^{\!\!\!\!-\frac{1}{2}}
\end{align*}
when $f(x)=c$ and 0 otherwise. Using Lemma 2 of~\cite{hayashi_general_2003} we can ``unravel'' these to obtain  
\begin{align}
\mathbbm{1}-\Lambda_{x;c,f}\leq 2(\mathbbm{1}-\Pi_x)+4\sum_{x'\neq x}\delta_{f(x'),c}\Pi_{x'}.
\end{align}
Next, consider the error probability for a given $x$ and $f$, and therefore $c=f(x)$:
\begin{align*}
p_{\rm err}(x,f)&=\tr{(\mathbbm{1}-\Lambda_{x;f(x),f})\varphi_x}\\
&\leq 2\tr{(\mathbbm{1}{-}\Pi_x)\varphi_x}+4\!\sum_{x'\neq x}\!\!\delta_{f(x'),f(x)}\tr{\Pi_{x'}\varphi_x}.
\end{align*}
Averaging over $f$ and using the 2-universal property simplifies the second term:
\begin{align*}
\overline{p}_{\rm err}&\leq 2\tr{\varphi_x(\mathbbm{1}-\Pi_x)}\\
&\phantom{\leq}+4\sum_{x'\neq x}{\rm Pr}_f[f(x')=f(x)]\tr{\varphi_x\Pi_{x'}}\\
&\leq 2\tr{\varphi_x(\mathbbm{1}-\Pi_x)}+4\cdot 2^{-m}\sum_{x'\neq x}\tr{\varphi_x\Pi_{x'}}\\
&\leq 2\tr{\varphi_x(\mathbbm{1}-\Pi_x)}+4\cdot 2^{-m}\sum_{x'\in\mathcal{X}}\tr{\varphi_x\Pi_{x'}}.
\end{align*}
Now average over $x$ to get
\begin{align*}
\overline{p}_{\rm err}&\leq 2\sum_{x\in\mathcal{X}}p_x\tr{{\varphi}_x(\mathbbm{1}-\Pi_x)}+4\cdot 2^{-m}\sum_{x'\in\mathcal{X}}\tr{\varphi\Pi_{x'}}.
\end{align*}
Using the form of $P^{XB}$ completes the proof.
\end{IEEEproof}

The second lemma is a corollary of a result proven in the context of hypothesis testing by Audenaert \emph{et al.}~\cite{audenaert_discriminating_2007,audenaert_asymptotic_2008} (in particular, see Eq.\ 24 of~\cite{audenaert_asymptotic_2008}). Here $\bpart{A}{+}$ denotes the projector onto the support of the positive part of $A$ and $\bpart{A}{-}$ the nonpositive part. 
\begin{lemma}
\label{lem:auden}
For $\rho,\sigma\geq 0$ and any $0\leq s\leq 1$,
\begin{align}
\tr{\rho\bpart{\rho-\sigma}{-}+\sigma\bpart{\rho-\sigma}{+}}\leq\tr{\rho^s\sigma^{1{-}s}}.
\end{align}
\end{lemma}

\begin{IEEEproof}[Proof of the Direct Part of Theorem~\ref{thm:csi}]
\noindent Let $P^{XB}=\bpart{\psi^{XB}-2^{-(m-1)}\mathbbm{1}^X\otimes \varphi^B}{+}$. Combining Lemma~\ref{lem:unravel} and  Lemma~\ref{lem:auden} with $s=\frac 12$, the bound on the error probability becomes 
\begin{align}
\overline{p}_{\rm err}&\leq \sqrt{8\cdot 2^{-m}}\,\tr{\sqrt{\psi^{XB}}\sqrt{\mathbbm{1}^X\otimes\varphi^B}}\\
&\leq \sqrt{8\cdot 2^{-m}}\left\|\sqrt{\psi^{XB}}\sqrt{\mathbbm{1}^X\otimes\varphi^B}\right\|_1\\
&\leq \sqrt{8\cdot 2^{-m}}\,\max_{\sigma^B}\,F(\psi^{XB},\mathbbm{1}^X\otimes\sigma^B)\\
&=\sqrt{8\cdot 2^{-(m-\hmax{}(X|B)_\psi)}}.
\end{align}
The second inequality 
is an immediate consequence of an alternate expression for the trace distance, $\|A\|_1=\max_U |\tr{UA}|$ for unitary $U$, which can be seen by using the polar decomposition $A=V\sqrt{A^\dagger A}$.
Choosing $m=\lceil\hmax{}(X|B)_\psi+2\log\frac{1}{\epsilon}\rceil+3$ then implies $\overline{p}_{\rm err}\leq \epsilon$. 

Now consider constructing a protocol for a nearby state $\bar{\psi}\in B_{\epsilon_1}(\psi)$ and suppose it achieves an error probability $\epsilon_2$. Since the trace distance is upper bounded by the purification distance, the error probability achieved by the protocol when applied to $\psi$ itself will not be more than $\epsilon_1+\epsilon_2$. Choosing a cq $\bar{\psi}$ minimizing the max-entropy, which can be done by virtue of Lemma~\ref{lem:cqsmooth} in the Appendix, it follows that we can set  
\begin{align}
m=\lceil\hmax{\epsilon_1}(X|B)_\psi+2\log\tfrac{1}{\epsilon_2}\rceil+3
\end{align}
and achieve this error probability. 
Since an error rate of $\epsilon=\epsilon_1+\epsilon_2$ can be achieved by selecting a hash function $f$ at random, there must exist one such function whose error rate does not exceed $\epsilon$. Finally, using $\lceil x\rceil\leq x+1$ completes the proof. 
\end{IEEEproof}

\begin{IEEEproof}[Proof of the Converse of Theorem~\ref{thm:csi}]
The converse rests on the fact that the max-entropy of $X$ given $BC$ must be small if $X$ is recoverable from $B$ and $C$. In fact, $p_{\rm err}\leq \epsilon$ implies $\hmax{\sqrt{2\epsilon}}(X|BC)_\psi\leq 0$.
To see this, suppose we apply the protocol generating the guess $X'$ of $X$ from $BC$. This is a quantum operation, and therefore the max-entropy cannot decrease (by Theorem 18 of~\cite{tomamichel_duality_2010}), meaning $\hmax{\epsilon}(X|BC)_\psi\leq \hmax{\epsilon}(X|X')_{\psi'}$. But since $p_{\rm err}\leq \epsilon$, it follows that the ideal output must be an element of $B_{\sqrt{2\epsilon}}(({\psi'})^{XX'})$ by the bounds between trace and purification distances~\cite{tomamichel_duality_2010}. Thus, $\hmax{\sqrt{2\epsilon}}(X|X')_{\psi'}=0$. 

Now select $\bar{\psi}^{XBC}\in B_{\sqrt{2\epsilon}}(\psi^{XBC})$ to minimize $\hmax{\sqrt{2\epsilon}}(X|BC)_\psi$. By the chain rule $\hmax{}(X|BC)_{\bar{\psi}}\geq \hmax{}(X|B)_{\bar{\psi}}-\log|C|$ of Lemma~\ref{lem:chain} in the Appendix, we have
\begin{align}
\log |C|\geq \hmax{}(X|B)_{\bar{\psi}}\geq \hmax{\sqrt{2\epsilon}}(X|B)_\psi,
\end{align}
completing the proof.
\end{IEEEproof}

\subsection{Common Randomness and Secret-Key Distillation}
By combining the data compression result with known results on randomness extraction and privacy amplification, we can easily construct one-shot protocols for distilling common randomness or secret-keys. 

Recall from~\cite{renner_security_2005,renner_universally_2005,koenig_sampling_2007,tomamichel_leftover_2011} that privacy amplification of the random variable $X$ against an adversary holding a possibly quantum register $E$ can yield a number of $\epsilon$-good random bits $\lext{\epsilon}(X|E)_\psi$ in accordance with the following bounds, where $\psi^{XE}=\sum_x p_x \ketbra{x}^X\otimes \varphi^E_x$. Note that previous results used the trace distance to define the smoothing, which accounts for the slight difference in the form of the upper bound given here. 
\begin{theorem}[Privacy Amplification~\cite{renner_security_2005,renner_universally_2005,koenig_sampling_2007,tomamichel_leftover_2011}]
For any $\epsilon_1+\epsilon_2=\epsilon$,\footnote{The extra +1 in the lower bound comes from rounding and the fact that~\cite{renner_security_2005} uses a slightly different distance measure.}
\label{thm:pa}
\begin{align*}
\hmin{\epsilon_1}(X|E)_\psi-2\log\tfrac{1}{\epsilon_2}\!+\!1\leq \lext{\epsilon}(X|E)_\psi\leq \hmin{\sqrt{2\epsilon}}(X|E)_\psi.
\end{align*}
\end{theorem}

To distill a secret key from a state $\psi^{XBE}=\sum_x p_x \ketbra{x}^X\otimes \varphi^{BE}_x$, in principle Alice and Bob need only first run the data compression scheme and then perform privacy amplification as in Theorem~\ref{thm:pa}. If they require the result to be uncorrelated with the classical message $C$, this can be simply lumped together with $E$ as defining the adversary. 
For such a two-step protocol, the overall approximation parameter will consist of a sum of the parameters of the various parts, by the triangle inequality of the trace distance. In the following, $\epsilon$ will denote the error in data compression, $\epsilon'$ the error in privacy amplification.

\begin{IEEEproof}[Proof of the direct part of Theorem~\ref{thm:skd}]
To prove the lower bound of Theorem~\ref{thm:skd}, we start by ignoring the supremum over functions taking $X$ to $(U,V)$. From Theorem~\ref{thm:pa} we have
\begin{align}
\ell_{\rm secr}^{\epsilon+\epsilon'}(X{;}B|E)_\psi\geq \hmin{{\epsilon'_1}}(X|CE)_\psi-2\log\tfrac{1}{\epsilon'_2}+1
\end{align}
Now we may simplify the righthand side by using Lemma~\ref{lem:minchain} of the Appendix (a slight modification of the chain rule part of Theorem 3.2.12 of~\cite{renner_security_2005}), which in the present context translates to $\hmin{\epsilon}(XC|E)\leq\hmin{\epsilon}(X|CE)+\log|C|$. Then, since $C$ is a deterministic function of $X$, it follows that $\hmin{\epsilon}(XC|E)=\hmin{\epsilon}(X|E)$. On the other hand, from Theorem~\ref{thm:csi} we have $\log|C|\leq  \hmax{{\epsilon_1}}(X|B)_\psi+2\log\frac{1}{\epsilon_2}+4$, meaning
\begin{align}
\ell_{\rm secr}^{\epsilon+\epsilon'}(X{;}B|E)_\psi
&\geq 
\hmin{{\epsilon'_1}}(X|E)_\psi\!-\!\hmax{{\epsilon_1}}(X|B)_\psi\nonumber\\
&\phantom{\geq}\,-2\log\tfrac{1}{\epsilon_2\, \epsilon'_2}-3.
\end{align}

Finally, the bound can be immediatly improved by considering preprocessing in which Alice first computes $U$ and $V$ from $X$, and publicly distributes $V$ to Bob (meaning Eve also obtains a copy). This yields 
\begin{align}
\ell_{\rm secr}^{\epsilon+\epsilon'}(X{;}B|E)\geq \sup_{(U,V)\leftarrow X}&\hmin{{\epsilon'_1}}(U|EV)_\psi\!-\!\hmax{{\epsilon_1}}(U|BV)_\psi\nonumber\\
&\,-2\log\tfrac{1}{\epsilon_2\,\epsilon'_2}-3.
\end{align}
Choosing $\epsilon_2'=\epsilon_2$ completes the proof.
\end{IEEEproof}

\begin{IEEEproof}[Proof of the converse of Theorem~\ref{thm:skd}]
Now we consider the converse for a generic key distillation protocol in which Alice generates $U$ and $V$ from $X$, broadcasts the latter and uses the former as the key, while Bob generates his version of the key $U'$ from $B$ and $V$. Let the cq state $\psi^{XBE}$ be the input to this process and $(\psi')^{UU'VE}$ be the output and suppose that the latter is an $\eps$-good approximation to a secret key of size $n=\ell_{\rm secr}^\eps(X;B|E)_\psi$ bits. As in the proof of the converse to Theorem~\ref{thm:csi}, this implies $\hmax{\sqrt{2\eps}}(U|U'V)_{\psi'}\leq 0$, and similar reasoning implies $\hmin{\sqrt{2\eps}}(U|EV)^\psi\geq n$. Thus,
\begin{align}
n&\leq \hmin{\sqrt{2\eps}}(U|EV)-\hmax{\sqrt{2\eps}}(U|U'V)\\
&\leq \hmin{\sqrt{2\eps}}(U|EV)-\hmax{\sqrt{2\eps}}(U|BV)\\
&\leq \sup_{(U,V)\leftarrow X}\left[\hmin{\sqrt{2\eps}}(U|EV)-\hmax{\sqrt{2\eps}}(U|BV)\right].
\end{align}
Here the second inequality follows from the non-decrease of the max-entropy under quantum operations (the data processing inequality), Theorem 18 of~\cite{tomamichel_duality_2010}.
\end{IEEEproof}

\section{Conclusions}

By characterizing the one-shot capabilities of data compression and secret key distillation in terms of smooth min- and max-entropies, we provide further evidence that a useful general theory of one-shot protocols does indeed exist and does not require the definition and study of new quantities for each individual protocol. 

Our results may also be specialized to the case of asymptotically-many \emph{non}-i.i.d.\ resources, and we find expressions for data compression and secret-key distillation in terms of spectral entropy rates, as introduced by Han \& Verd\'u~\cite{verdu_general_1994,han_information-spectrum_2002} for the classical case and generalized by Hayashi \& Nagaoka~\cite{hayashi_general_2003} to quantum information. For instance, inserting the result of Theorem~\ref{thm:csi} into the correspondence formulas derived in~\cite{datta_smooth_2009}, we immediately find an expression for optimal data compression in terms of a spectral entropy rate, which complements a result on data compression derived in~\cite{bowen_quantum_2006} (the latter applies to a setting without side information, but where the data to be compressed is quantum-mechanical). Furthermore, we note that the known expressions for data compression and secret-key distillation in the i.i.d.\ case can be readily recovered from our results by virtue of the Quantum Asymptotic Equipartition Property~\cite{tomamichel_fully_2009}. 

We may also immediately infer two entropy relations from our results, a chain rule for max-entropies and an uncertainty relation similar to the one given in~\cite{berta_uncertainty_2010}. To derive a chain rule, consider the problem of compressing a joint classical random variable $XY$, with quantum side information available at the decoder.  One way to construct a compression protocol for $XY$ is to first compress $X$ in a way suitable for a decoder with access to $B$ and then to compress $Y$ for a decoder with access to $XB$. Clearly this will not be better than the optimal protocol. 

Suppose that the first step succeeds in identifying $X$ with average error probability $\eps_x$. We may imagine that the decoder coherently performs the appropriate measurement on $B$ and stores the result in an auxiliary system $X'$; using  the purification of the input cq state $\psi^{XB}$ it is then easy to show that the actual result of this process, the state $\xi^{XX'B}$ is essentially the same as the ideal output $\widetilde{\xi}^{XX'B}$ in which $X'$ is simply a copy of $X$ and the side information $B$ is untouched. In particular, including the random variable $Y$, we have $\frac12\left\|\xi^{XX'YB}-\widetilde{\xi}^{XX'YB}\right\|_1\leq \sqrt{2\eps_x}$. 

Since the side information in $B$ is essentially unchanged, it can subsequently be used to help determining $Y$. Now let $\eps_y$ be the average error probability for a compression scheme of $Y$ given an exact copy of $X$ at the decompressor, i.e.\ the input described by $\widetilde{\xi}^{XX'YB}$. Using $Y'$ to store the output of the measurement, the triangle inequality and contractivity of the trace distance under partial trace implies $\tfrac12\left\|\xi^{XX'YY'}-\widetilde{\xi}^{XX'YY'}\right\|_1\leq \sqrt{2\eps_x}+\eps_y$, where $\widetilde{\xi}^{XX'YY'B}$ is again the ideal output in which $X'=X$ and $Y'=Y$. Working out the trace distance for states of this form reveals that it simply equals the error probability, so the total probability of incorrectly determining $X$ and $Y$ is no greater than $\sqrt{2\eps_x}+\eps_y$. After setting $\eps_x=\eps_y=2\eps$ and $\eps'=2(\eps+\sqrt{\eps})$, 
Theorem~\ref{thm:csi} implies
\begin{align}
\hmax{\eps}(X|B)_\psi+\hmax{\eps}(Y|XB)_\psi&\geq 
 \hmax{\sqrt{2\eps'}}(XY|B)_\psi\nonumber\\
 &\phantom{\geq}-4\log\tfrac1\eps-8.
\end{align}

Another immediate application of our results is the derivation of an uncertainty relation similar to the one given in~\cite{berta_uncertainty_2010}. In contrast to the proof in~\cite{berta_uncertainty_2010}, the derivation makes use of the operational meaning of these quantities. A recent result by one of us shows that, just as min- and max-entropy are in some sense dual, protocols for privacy amplification and data compression are dual, too~\cite{renes_duality_2011}. Specifically, a linear protocol for data compression with side information can be transformed into a linear protocol for privacy amplification, and, under certain conditions, vice versa. Thus, we can start with a data compression protocol operating in accord with Theorem~\ref{thm:csi} and transform it into a privacy amplification protocol, at which point it is subject to the constraints of Theorem~\ref{thm:pa}. The end result of this analysis, carried out in more detail in~\cite{renes_duality_2011}, is the following uncertainty relation, valid for arbitrary $\epsilon>0$:
\begin{align}
\label{eq:unc}
\hmin{\varepsilon}(X^A|R)_\psi+\hmax{\varepsilon}(Z^A|B)_\psi\geq \log_2 d-8\log\tfrac{1}{\varepsilon}-12.
\end{align}
Here $A$ is a system of dimension $d$, held by Alice. She can perform one of two measurements, corresponding to the two bases which are eigenbases of the operators $X^A=\sum_{k=0}^{d{-}1}\ket{k{+}1}\bra{k}$ and $Z^A=\sum_{k=0}^{d{-}1} \omega^k\ketbra{k}$, for $\omega=e^{2\pi i/d}$. $B$ and $R$ are additional (quantum) systems whose purpose is to help predict the outcomes of hypothetical $Z^A$ and $X^A$ measurements, respectively; the capability of one constrains the capability of the other. 

We have only concentrated on protocols whose ultimate aim is to process classical information, albeit perhaps stored in quantum states; protocols manipulating quantum information, such as in entanglement distillation, are not considered. However,  there exists a strong connection between the two, at least in the asymptotic i.i.d.\ scenario, and it could be fruitful to extend this connection to the one-shot case. 

Taking the case of entanglement distillation, the first proof of the achievable distillation rate proceeds by first establishing the achievable rate of secret-key distillation and then showing that coherently performing the protocol results in an entanglement distillation protocol~\cite{devetak_distillation_2005}. An equivalent distillation protocol can be constructed by combining two protocols for classical data compression with quantum side information, one for each of two complementary bases (related by Fourier transform), as shown in~\cite{renes_physical_2008}. Thus, there are two possible ways to construct one-shot entanglement distillation from the results presented here, and it would be interesting to compare with more ``fully'' quantum approaches, such as~\cite{berta_single-shot_2008,buscemi_distilling_2010}.

\appendices

\section{CQ Smoothing}
In this appendix we show that the optimal state for smoothing the max-entropy of a cq state is itself a cq state. A similar result was shown for the min-entropy in Remark 3.2.4 of~\cite{renner_security_2005}.
\begin{lemma}
\label{lem:cqsmooth}
  Let $\rho^{X B}$ be a cq state and let $\epsilon \geq 0$. Then there exists a cq state $\bar{\rho}^{X B} \in B_{\epsilon}(\rho^{X B})$ such that
  \begin{align*}
    \hmax{\epsilon}(X|B)_{\rho} = \hmax{}(X|B)_{\bar{\rho}}
  \end{align*}
\end{lemma}

\begin{IEEEproof}
  Observe first that a state $\rho^{X B}$ is a cq state if and only if it has a purification  $\rho^{X X' B C} = \ketbra{\Psi}$ of the form
  \begin{align} \label{eq:cqdual}
    \ket{\Psi}^{X X' B C} = \sum_x \alpha_x \ket{x}^X \otimes \ket{x}^{X'} \otimes \ket{\phi_x}^{B C}
  \end{align}
  where $X$ and $X'$ are isomorphic and where $\{\ket{x}\}_x$ is an orthonormal basis of these spaces.

  By the duality between smooth min- and max-entropy~\cite{tomamichel_duality_2010}, it suffices to show that there exists a (subnormalized) vector $\ket{\bar{\Psi}}$ of the form~\eqref{eq:cqdual} such that $\bar{\rho}^{X X' B C} = \ketbra{\bar{\Psi}} \in B_\eps(\rho^{X X' B C})$ and 
  \begin{align} \label{eq:smoothminv}
    \hmin{\epsilon}(X| X' C)_{\rho} = \hmin{}(X|X' C)_{\bar{\rho}} \ .
  \end{align}
  To show this, let $\hat{\rho}^{X X' C} \in B_\epsilon(\rho^{X X' C})$ such that the min-entropy is maximized, i.e., 
  \begin{align*}
    \hmin{\epsilon}(X | X' C)_{\rho} = \hmin{}(X|X' C)_{\hat{\rho}} \ .
  \end{align*}
  By the definition of the purified distance, there exists a purification $\hat{\rho}^{X X' B C} = \ketbra{\hat{\Psi}}$ that is $\epsilon$-close to $\rho^{X X' BC}$, i.e., 
 $
    |\bracket{\hat{\Psi}}{\Psi}| = \sqrt{1-\epsilon^2}.
  $
  Using the projector $P^{X X'} \equiv \sum_{x} \ketbra{x}^{X} \otimes \ketbra{x}^{X'}$, define $\ket{\bar{\Psi}} \equiv (P^{X X'} \otimes \mathbbm{1}^{B C}) \ket{\hat{\Psi}}$. Since $(P^{X X'} \otimes \mathbbm{1}^{B C}) \ket{\Psi} = \ket{\Psi}$, we have
  \begin{align*}
    |\bracket{\bar{\Psi}}{\Psi}|
  =
    |\bra{\hat{\Psi}} (P^{X X'} \otimes \mathbbm{1}^{B C}) \ket{\Psi}|
  = 
    |\bracket{\hat{\Psi}}{\Psi}|
  = 
    \sqrt{1-\epsilon^2} 
  \end{align*}
  and, hence, $\bar{\rho}^{X X' B C} \in B_{\epsilon}(\rho^{X X' B C})$. 

  By the definition of min-entropy, there exists a state $\sigma^{X' C}$ such that
  \begin{align*}
    \hat{\rho}^{X X' C} \leq \lambda \mathbbm{1}^{X} \otimes \sigma^{X' C} 
  \end{align*}  
  for $\lambda = 2^{-\hmin{}(X | X' C)_{\hat{\rho}}}$.  Applying the projection $P^{X X'}$ on both sides of this operator inequality gives
  \begin{multline*}
    \bar{\rho}_{X X' C} = (P^{X X'} \otimes \mathbbm{1}^{C}) \hat{\rho}^{X X' C} (P^{X X'} \otimes \mathbbm{1}^{C}) \\
  \leq 
    \lambda (P^{X X'} \otimes \mathbbm{1}^{C}) (\mathbbm{1}^{X} \otimes \sigma^{X' C}) (P^{X X'} \otimes \mathbbm{1}^{C})
  \leq
    \lambda \mathbbm{1}^{X} \otimes \bar{\sigma}^{X' C}
  \end{multline*}
  for $\bar{\sigma}^{X' C} \equiv \sum_x (\ketbra{x} \otimes \mathbbm{1}^C) \sigma^{X' C} (\ketbra{x} \otimes \mathbbm{1}^C)$. This immediately implies that
  \begin{align*}
     \hmin{\epsilon}(X| X' C)_{\rho} \leq \hmin{}(X|X' C)_{\bar{\rho}} 
  \end{align*}
  Since the opposite inequality ($\geq$) holds by definition of the smooth min-entropy, we have proved~\eqref{eq:smoothminv}, which concludes the proof.
 \end{IEEEproof}

\section{Chain Rules}

Here we prove two chain rules which are important for the converses of Theorems~\ref{thm:csi} and~\ref{thm:skd}.
\begin{lemma}
\label{lem:chain}
For $C$ classical, 
\begin{align}
\hmax{}(A|BC)&\geq \hmax{}(A|B)-\log|C|.
\end{align}
\end{lemma}
\begin{IEEEproof}
The general form of the state is $\rho^{ABC}=\sum_c{p_c}\rho^{AB}_c\otimes \ketbra{c}^C$, where $p_c$ is a probability distribution and the $\rho_c^{AB}$ are normalized states. A purification of $\rho^{ABC}$ is $\ket{\psi}^{ABCRC'}=\sum_c\sqrt{p_c}\ket{c,c}^{CC'}\ket{\psi_c}^{ABR}$, for $\ket{\psi_c}^{ABR}$ a purification of $\rho_c^{AB}$. By duality, Eq.~\ref{eq:dual}, the stated inequality is equivalent to $\hmin{}(A|RCC')_\psi\geq \hmin{}(A|RC')_\psi-\log|C'|$ (since $|C|=|C'|$). We now establish the equivalent form. 

First, make the definitions $\ket{\varphi_c}^{ABCR}=\sqrt{p_c}\ket{c}^{C}\ket{\psi_c}^{ABR}$, $\ket{\varphi}^{ABCR}=\sum_c \ket{\psi_c}^{ABCR}$, and $\widetilde{\varphi}^{ABCRC'}=\sum_c \ketbra{c}^{C'}\otimes\varphi_c^{ABCR}$ and let   $\hmin{}(\rho^{AB}|\sigma^B)$ be the min-entropy as defined in Eq.~\ref{eq:hmin}, but without the maximization over $\sigma^{B}$.
Using Lemma 3.1.14 of~\cite{renner_security_2005} we conclude 
\begin{align*}
\hmin{}(\varphi^{ARC}|\sigma^{RC})&\geq \hmin{}(\widetilde{\varphi}^{ARCC'}|{\sigma}^{RCC'})-\hmax{}(\widetilde{\varphi}^{C'})\\
&\geq  \hmin{}(\widetilde{\varphi}^{ARCC'}|{\sigma}^{RCC'})-\log|C|,
\end{align*}
for arbitrary $\sigma^{RCC'}$ and ${\sigma}^{RC}={\rm Tr}_{C'}[\sigma^{RCC'}]$, and 
where the second line follows from the fact that the max-entropy is upper bounded by the logarithm of the state-space dimension (alphabet size). If we choose $\bar{\sigma}^{RCC'}$ so that $\hmin{}(\widetilde{\varphi}^{ARCC'}|\bar{\sigma}^{RCC'})=\hmin{}(A|RCC')_{\widetilde{\varphi}}$ then we obtain
\begin{align*}
\hmin{}(A|RC)_\varphi &\geq \hmin{}(\varphi^{ARC}|\sigma^{RC})\\
&\geq \hmin{}(A|RCC')_{\widetilde{\varphi}}-\log|C|.
\end{align*}

Now observe that unitarily copying $C$ to $C'$ in $\varphi^{ABCR}$ results in $\psi^{ABCRC'}$. This will not affect the conditional entropy, so $\hmin{}(A|RCC')_\psi=\hmin{}(A|RC)_\varphi$. Likewise, $C$ can be deleted from $C'$ in $\widetilde{\varphi}^{ABCRC'}$, producing the state $\psi^{ABRC'}=\sum_c p_c \ketbra{c}^{C'}\otimes \psi_c^{ABR}$.  Thus, $\hmin{}(A|RC')_\psi=\hmin{}(A|RCC')_{\widetilde{\varphi}}$, completing the proof.
\end{IEEEproof}

\begin{lemma}
\label{lem:minchain}
$\hmin{\epsilon}(AB|C)\leq\hmin{\epsilon}(A|BC)+\log|B|$.
\end{lemma}
\begin{IEEEproof}
Start by choosing $\bar{\psi}^{ABC}\in B_\epsilon(\psi^{ABC})$ such that $\hmin{\epsilon}(AB|C)_\psi=\hmin{}(AB|C)_{\bar{\psi}}$. From the definition of conditional min-entropy, we have $\bar{\psi}^{ABC}\leq 2^{-\hmin{}(AB|C)_{\bar{\psi}}}\mathbbm{1}^{AB}\otimes\sigma^C$, for the optimal $\sigma^C$. Defining $\eta^{BC}=\frac{1}{|B|}\mathbbm{1}^B\otimes\sigma^C$, this is equivalent to $\bar{\psi}^{ABC}\leq {2^{-\hmin{}(AB|C)_{\bar{\psi}}}|B|\mathbbm{1}^{A}\otimes\eta^{BC}}$. Using the definition once again, we can easily see that $2^{-\hmin{}(AB|C)_{\bar{\psi}}}|B|\geq 2^{-\hmin{}(A|BC)_{\bar{\psi}}}$, or equivalently, $\hmin{}(AB|C)_{\bar{\psi}}\leq \hmin{}(A|BC)_{\bar{\psi}}+\log|B|$. Finally, the fact that $\hmin{}(A|BC)_{\bar{\psi}}\leq \hmin{\epsilon}(A|BC)_\psi$ completes the proof.
\end{IEEEproof}

\section*{Acknowledgments}
The authors are grateful to Mark M.\ Wilde and Marco Tomamichel for careful reading of the manuscript.  
JMR acknowledges the support of CASED (www.cased.de). RR acknowledges support from the Swiss National Science Foundation (grant Nos.~200021-119868 and 200020-135048) as well as the European Research Council (ERC) (grant No~258932).

\bibliographystyle{IEEEtran}
\bibliography{csidist}

\end{document}